# Selective Defect Formation in Hexagonal Boron Nitride

Irfan H. Abidi[1,2,†], Noah Mendelson[3,†], Toan Trong Tran[3], Abhishek Tyagi[1], Minghao Zhuang[1], Lu-Tao Weng[1,4], Barbaros Özyilmaz[2,5], Igor Aharonovich[3, *], Milos Toth[3, *], and Zhengtang Luo[1, *]

[1] Department of Chemical and Biological Engineering, William Mong Institute of Nano Science and Technology and Hong Kong Branch of Chinese National Engineering Research Center for Tissue Restoration and Reconstruction, The Hong Kong University of Science and Technology, Clear Water Bay, Kowloon, 999077, Hong Kong

[2] Centre for Advanced 2D Materials, National University of Singapore, 117542, Singapore

[3] School of Mathematical and Physical Sciences, University of Technology Sydney, Ultimo, New South Wales 2007, Australia

[4] Materials Characterization and Preparation Facility, The Hong Kong University of Science and Technology, Clear Water Bay, Kowloon, 999077, Hong Kong

[5] Department of Physics, National University of Singapore, 117551, Singapore

† These authors contributed equally to this work

*Igor.Aharonovich@uts.edu.au, *Milos.Toth@uts.edu.au, *keztluo@ust.hk

**Abstract:** Luminescent defect-centers in hexagonal boron nitride (hBN) have emerged as a promising 2D-source of single photon emitters (SPEs) due to their high brightness and robust operation at room temperature. The ability to create such emitters with well-defined optical properties is a cornerstone towards their integration into on-chip photonic architectures. Here, we report an effective approach to fabricate hBN single photon emitters (SPEs) with desired emission properties in two isolated spectral regions *via* the manipulation of boron diffusion through copper during atmospheric pressure chemical vapor deposition (APCVD)—a process we term gettering. Using the gettering technique we deterministically place the resulting zero-phonon line (ZPL) between the regions 550-600 nm or from 600-650 nm, paving the way for hBN SPEs with tailored emission properties across a broad spectral range. Our ability to control defect formation during hBN growth provides a simple and cost effective means to improve the crystallinity of CVD hBN films, and lower defect density making it applicable to hBN growth for a wide-range of applications. Our results are important to understand defect formation of quantum emitters in hBN and deploy them for scalable photonic technologies.

Robust and photostable single photon emitters (SPEs) underpin a number of promising quantum information science and technologies.[1-4] Solid-state quantum light sources are highly sought after for their ease of integration into on-chip architectures, and rapid progress has been made recently with a variety of solid-state sources such as quantum dots,[5, 6] color centers in diamond,[7, 8] silicon carbide,[9, 10] rare-earth materials,[11] carbon nanotubes,[12] and layered van der Waals materials.[3] One of the most promising candidates are quantum emitters in hexagonal boron nitride (hBN), which have recently emerged as a robust solid-state platform capable of hosting bright,[13-17] linearly polarized,[13, 18] and optically stable SPEs operating at room temperature with high photon purity.[19, 20] While the zero-phonon lines (ZPLs) of hBN quantum emitters have been known to display a wide spread of energies (~1.6-2.4eV), implying the existence of multiple defect species,[15, 18, 21-23] recent progress utilizing chemical vapor deposition (CVD) has shown that this spread can be reduced to ~100 meV.[24, 25] This reduced ZPL energy distribution is useful both for applications and for basic studies of the defects responsible for the emissions. However, to fully understand and exploit the diversity of emissions reported in hBN further advances in controlling the observed emission spectra and emitter density are required.

In order to target the incorporation of particular structural defects during CVD growth, a method to modify the dominant defect formation processes is required. *In-situ* monitoring of hBN growth on Cu has clarified that hBN growth is subject to complex interaction between the precursor species, and the catalyst surface/bulk.[26] As a result, controlling the interactions between the precursor species and the catalyst is critical to achieve highly crystalline hBN growth, but also offers a promising avenue to controlled defect engineering during CVD growth of hBN. In other material systems such as InAs/GaAs quantum dots modifying catalyst properties such as lattice mismatch during epitaxial growth has been definitively linked to defect creation.[27] Similarly graphene is known to be dependent of the interactions with the catalyst bulk/surface, where growth depends strongly on the relative phase of the catalyst and the supply of carbon in and out of the catalyst, controlling both defect formation and density.[28, 29] Despite added complications for CVD growth of hBN, especially due to differences in precursor solubility, inspiration can be gleaned for the catalytic growth of III-V nanowires, which also suffered from similar issues, yet can be controlled to yield atomically precise growth by doping the Au catalyst with a precise concentration of Gallium.[30]

In this work we modify the B diffusion into Cu through a process that we term backside boron gettering (BBG), to realize hBN of superior crystallinity compared to traditional pre-oxidation techniques alone. More importantly, we use this technique to control the incorporated defects in CVD hBN, allowing us to place the resulting SPE ZPLs within a specific spectral window of either 550-600 nm or 600-650 nm, with near exclusive incorporation (>85%) under each growth condition. These results also suggest the operation of two different structural defects within the isolated ZPL regions. The reliable positioning of the spectral properties surpasses that achieved with alternate top-down or bottom-up approaches,[31, 32] paving the way for hBN SPEs with tailored optical properties across a broad spectral range. Our work presents a template for the control of defect formation in hBN, yielding SPEs with predictable ZPL energies, through controlling the diffusion of B into copper during CVD.

**Results and Discussion:**

At the elevated temperatures, often used in CVD growth of hBN, diffusion of B and N species through the metallic catalyst play a critical role for layer-controlled growth and fabrication of high-quality hBN.[33-35] Due to the difference in solubility of B and N constituents in Cu, where B is readily incorporated into the bulk lattice while N experiences negligible incorporation, the interaction between the catalyst surface/bulk will dictate the supply of the precursor molecules and thus control the growth process.[26] Furthermore, growth past the monolayer occurs only at the interface of the hBN and Cu catalyst, so interaction with the bulk catalyst continuous to be the dominant factor in growth past 1 layer,[26] suggesting any rational control of defect formation should target modifying the properties of the Cu catalyst. As the properties of the Cu catalyst are directly tied to the absorption of B into the bulk, resulting in an expansion of the Cu lattice constant, and the leaching of adsorbed boron species back to the surface during the cooling process, modifying this B uptake is a promising route to control defect formation. Specifically, the diffusion of B through Cu can be mitigated by stacking the Cu foil on a getter substrate such as nickel (Ni),[36, 37] provided the solubility of boron is higher in Ni than in Cu.[38]

Figure 1a displays a schematic of the APCVD setup used for hBN growth on Cu foil (for more information see Figure S1 and S2). To investigate the role of the Cu support on hBN growth, the Cu foil was stacked on one of two support substrates during growth–quartz as an inert control substrate (Cu/Quartz) and nickel as a gettering substrate (Cu/Ni). Ammonia borane ($NH_3BH_3$) powder was used as a precursor for hBN growth, and was placed upstream in the CVD tube furnace. During growth, ammonia borane is decomposed at 100 °C to produce gaseous species such as aminoborane, borazine and hydrogen,[39, 40] which are carried upstream by a flow of $H_2$/Ar, and delivered to the Cu foil, kept at 1040 °C (*cf* methods). In addition to the use of pristine Cu, we also fabricated reference substrates in which the Cu surface was oxidized by annealing in ambient conditions at 200 °C for 5 minutes prior to hBN growth.[41] These are designated Cu(O)/quartz and Cu(O)/Nickel for substrates placed on quartz and Ni support substrates. The respective sample compositions are depicted in Figure S3, and are referred to hereafter as "standard" for Cu/Quartz, "pre-oxidized" for Cu(O)/Quartz, and "gettered" for Cu(O)/Nickel.

Figure 1b-g show optical images of the top and backside of the Cu foil after 10 minutes of CVD growth, highlighting the variation in growth kinetics from the three substrate types. Figure 1b and 1c reveal that for the standard and pre-oxidized growths, the Cu top-surface is almost completely covered with hBN film after 10 minutes of growth, while the hBN coverage is substantially lower for gettered growth conditions, Figure 1d. This variation in nucleation and growth rates result in different times to full hBN coverage on the top Cu surface, Figure S4. The substrate effects become even clearer when evaluating the backside of the Cu foil, Figure 1e-g. hBN growth on the backside of Cu is the result of diffusing precursor vapors at the interface of Cu foil and the support substrate, but displays slower growth kinetics limited by the confined space.[42] Figure 1f reveals the critical role of oxygen in further lowering the nucleation density of hBN domains, resulting in large size single-crystal domains ~10 times larger than observed without pre-oxidation.[41] However, the influence of the Ni getter is even more dramatic, as no hBN back-side growth is observed, Figure 1, even after a growth duration of 2 hours, Figure S5.

Furthermore, the effect of modifying the bulk catalyst properties are manifested on the top-side growth surface as well, as changes to the single crystal domain shape and a reduction in growth rate are observed, indicating a modification in the supply dynamics of B species for growth from the catalyst. This discrepancy between hBN growth on Cu/Quartz and Cu/Nickel is similar to that observed for carbon gettering growth of graphene.[36, 37] Therefore, our results indicate a similar gettering mechanism for boron atoms by the nickel support on the backside of the Cu foil during CVD growth of hBN. The appeal of this method is furthered by its ease of implementation, low cost, and the absence of dangerous chemicals such as are common for electropolishing.[43]

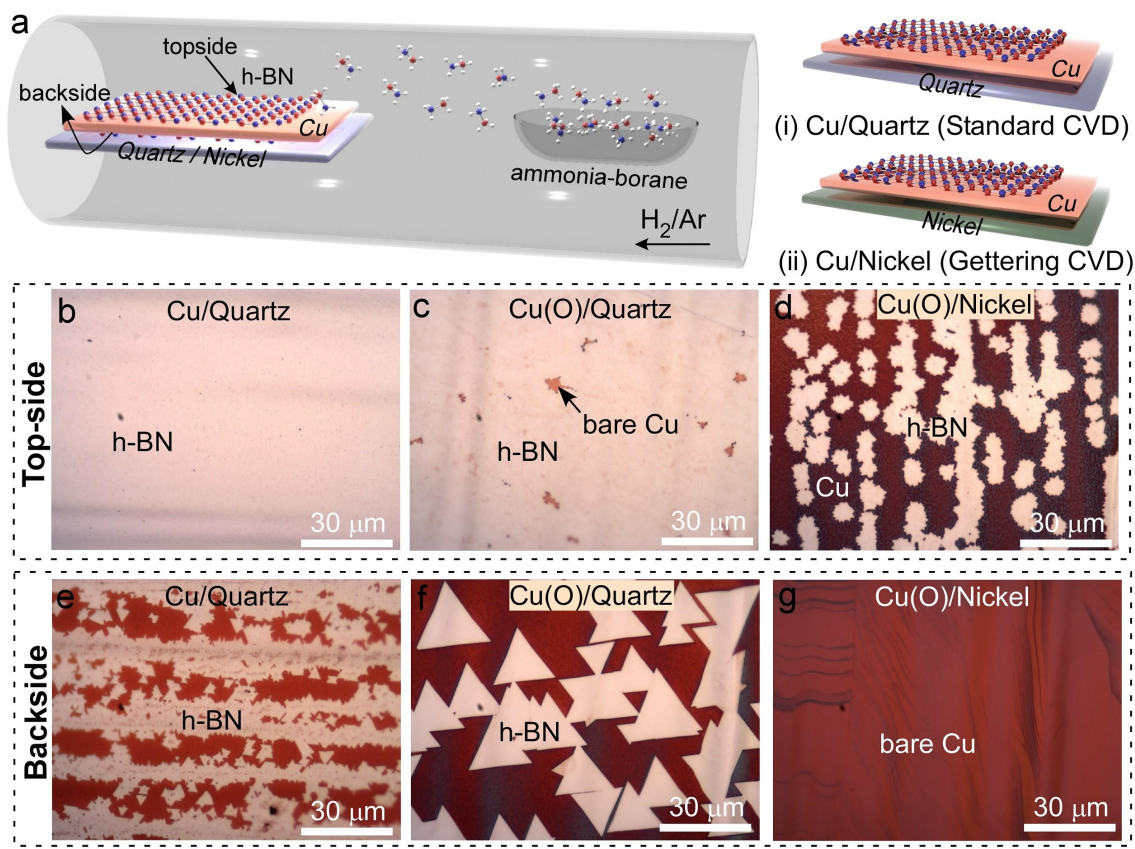

*Figure 1. APCVD growth of h-BN on Cu.* *(a) Schematic of APCVD of h-BN on Cu foil resting on quartz (Cu/Quartz) or nickel (Cu/Nickel) support substrate. The h-BN growth on top and backside of the Cu foil (indicated by arrows) is studied in this work. The optical images of as-grown h-BN on Cu top side (b-d) and backside (e-f) after 10 minutes of growth, is shown. The acronym Cu(O) represent the conditions where pre-oxidized Cu is used for the growth. On the Cu topside, for the same growth time duration, Cu(O)/Nickel (d) shows lower nucleation density as compared to Cu/Quartz (a,c), as h-BN fully covered the Cu surface in 10 minutes. On the Cu backside, the nucleation density decreases while using pre-oxidized Cu on quartz (f) resulting in large single-crystals of h-BN. In contrast, there is no h-BN growth for Cu/Nickel configuration (g).*

Despite the observed differences in growth kinetics, all three sample types were confirmed to produce high quality hBN thin-films on the top growth surface. Figure 2a shows the Raman spectra taken from the transferred hBN films onto 300 nm SiO$_2$/Si wafers. The peak at ~1369 cm$^{-1}$ is the characteristic Raman hBN E$_{2g}$ vibrational mode,[44, 45] while the peak at ~1450 cm$^{-1}$ results from the Si 3TO mode,[41] generated from the underlying substrate. The E$_{2g}$ modes were fit with a Lorentzian, giving full-width at half maximum (FWHM) of 24.35 cm$^{-1}$ for the standard grown hBN, which decreases to 17.60 cm$^{-1}$ and 15.90 cm$^{-1}$ for pre-oxidized and gettering growth conditions, respectively, Figure S6. This confirms the superior crystallinity of hBN grown films by the gettering method.[45, 46] The increasing crystallinity can be ascribed to modification of the Cu catalyst during gettering growth, which slows nucleation and growth kinetics, and reduces the effects of lattice constant changes, and high defect density growth during cooling by boron precipitation from the catalyst.[26, 33] We also observed a corresponding decrease in the roughness of the transferred hBN films *via* AFM, Figure S6, where R$_a$ values decreased from 1.29 to 1.09 to 0.82 respectively for the standard, pre-oxidized, and gettered samples, suggesting a smoother, more crystalline surface. The hBN film thickness for each film was found to be below 4nm, with measured values ~3.5nm, ~2nm and ~1.5nm, for the standard, pre-oxidized, and gettered growths respectively.

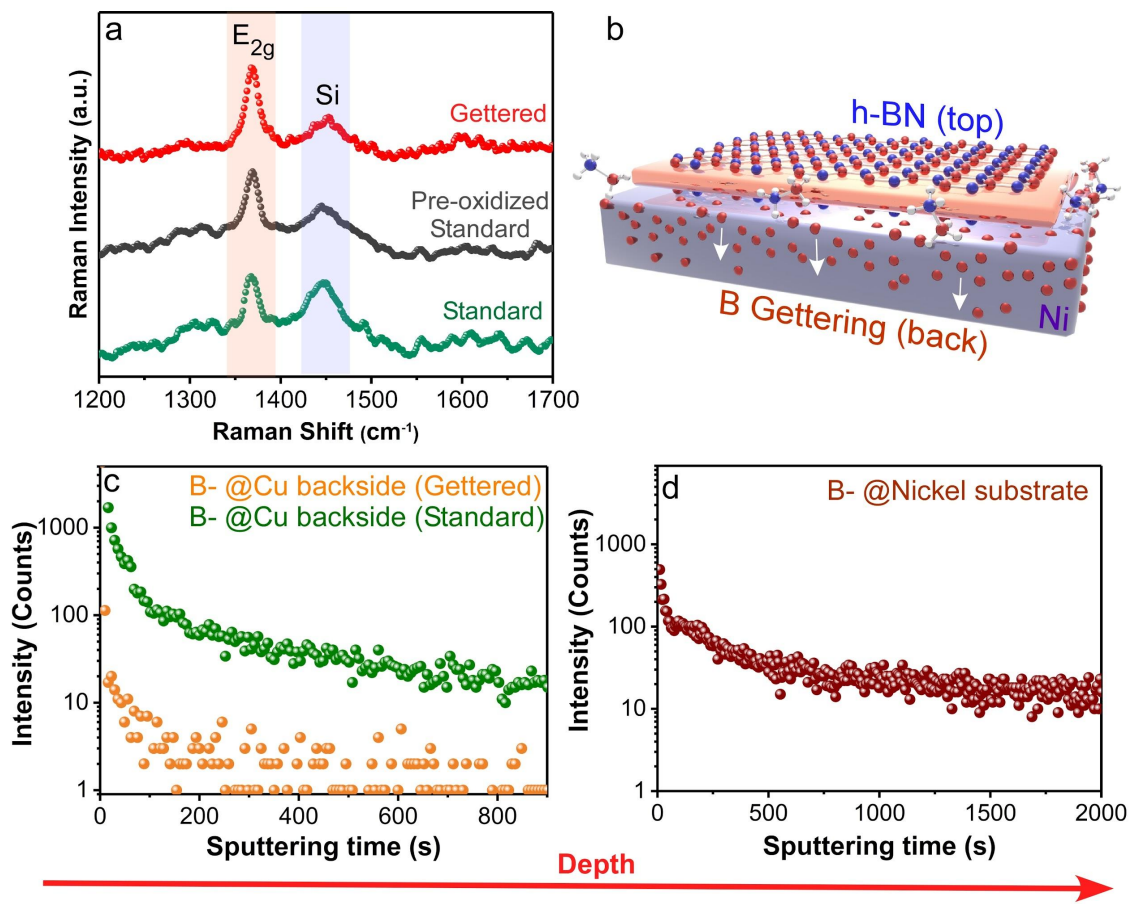

*Figure 2. Revealing the B Gettering Mechanism of Nickel Support Substrate. (a) Representative Raman spectra of h-BN films obtained by different growth conditions used in this work. (b) Schematic illustrating backside B gettering mechanism of h-BN growth on Cu. Nickel support is used as B getter substrate at backside of the Cu foil during CVD growth. (c) ToF-SIMS depth profile of B- ions from backside of the Cu foil after CVD growth, backed by nickel (getter growth) and quartz (standard growth). The higher concentration of B- for Cu/quartz indicates the B diffusion through Cu during standard CVD growth, while B diffusion is mitigated for gettering growth. (d) ToF-SIMS depth profile of B- ions within the nickel support after CVD growth. The higher concentration of B- deep inside nickel substrate reveals its role as gettering the B species at the Cu/Ni interface.*

To explain the increasing crystallinity of the hBN films, and the observation that no growth occurs on the backside of the Cu foil with Ni present, we propose a gettering mechanism, shown schematically in Figure 2b. Where given the higher solubility of B in Ni *versus* Cu,[38] it is expected that as precursor species accumulate at this interface, the B atoms preferentially diffuse into the Ni foil instead of into Cu or remaining on the back-side surface to catalyze hBN growth. To confirm the proposed BBG mechanism we analyzed the concentrations of B within the Ni and Cu foils post growth, and compared this with the Cu foil in the absence of Ni backing by time-of-flight secondary ion mass spectroscopy (ToF-SIMS) depth profile analysis, after the respective CVD growths. Figure 2c illustrates the B ion depth profile as a function of sputtering time within the bulk Cu from the back-side surface, *i.e.* that facing the quartz or nickel support. The higher concentration of B within the Cu/quartz specifies the B diffusion through the bulk Cu during standard CVD growth. However, the depth profiling of Cu backed with nickel shows the distribution of B only on the Cu surface, which diminishes rapidly with increasing depth, indicating the lack of B diffusion within the bulk Cu for gettering CVD growth. Moreover, the intense B signals generated by the post-growth depth profile of the nickel support, reveals the preferential B diffusion within nickel, Figure 2d, confirming nickel as an effective B gettering substrate. Note that N signals were below the detection limit of ToF-SIMS within Cu for all investigated growth conditions, which is attributed to the very low solubility of N in Cu at this growth temperature.[45, 47] Hence, our gettering approach enables us to mitigate B diffusion into Cu, improving the crystallinity of resulting hBN growth on the top-side Cu surface, and offering a controllable way to engineer defect formation during CVD growth.

This effective modulation of B diffusion, not only afford a facile means to improve crystal quality of the hBN, but has two important consequences relevant to defect creation in hBN. First, the modification of the Cu lattice constant, which expands due to the B diffusion,[40] which can induce strain into the growth process[48] and modify defect creation during epitaxial growth.[49] And second, modifying the Cu catalyst resulting in changes to the energetics of precursor decomposition and supply rates, as well as mitigating the resulting growth during the cooling phase of CVD growth as B is precipitated from the copper.[26, 33] To quantify the effects of these changes on defect creation, and specifically the creation of single photon emitting defects, we characterized the photoluminescence properties of quantum emitters incorporated in the as-grown hBN films using a home-built confocal microscope setup, and a 532-nm continuous-wave (CW) excitation laser (*cf.* methods).

We found that all three types of APCVD grown samples hosted luminescent defect centers, with the representative room-temperature spectra from each displayed in Figures 3a-c. To confirm the single photon nature of the emission, we recorded second-order autocorrelation measurements using a Hanbury-Brown-Twiss (HBT) interferometer for all three growth types, Figure 3d. The emission lines were filtered using a tunable bandpass filter (550-610 nm), and all three samples showed $g^2(\tau)<0.5$, confirming the quantum nature of the emission. The lowest values were obtained consistently from the Ni backed samples with $g^2(\tau)\leq 0.2$, while the standard growth and pre-oxidized samples typically around $g^2(\tau)=0.4$, and $g^2(\tau)=0.25$ respectively. Note none of the autocorrelation measurements in Figure 3d were background corrected. Observed SPE properties were found to be consistent across multiple growths attempts under each set of conditions.

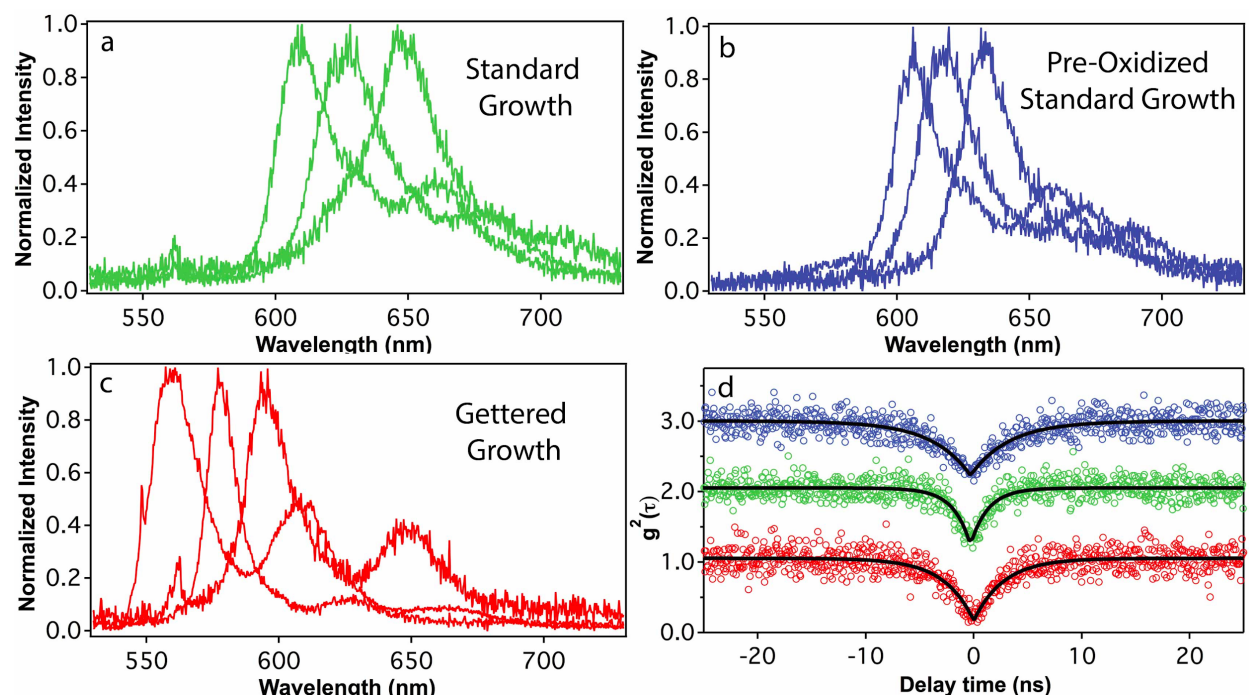

*Figure 3. Quantum Emission from APCVD hBN. (a) Representative spectra from three emitters acquired from samples grown on $SiO_2$ backed copper foil. (b) Representative spectra from three emitters acquired from samples grown on $SiO_2$ backed pre-oxidized copper foil. (c) Representative emitters grown on Ni backed copper foil. (d) Autocorrelation measurements from all three sample types, denoted by color coding to match spectra in a-c offset for clarity, displaying the quantum nature of the emission from each representative sample type.*

The spectra of 76 individual emitters were collected for each sample type, and all three types of APCVD materials displayed ZPLs almost exclusively occupying a narrow spectral window ≤50 nm. For standard growth samples, 100% of the 76 investigated emitters displayed ZPL positions from 600-650 nm, Figure 4a, hinting at the incorporation of a single structural defect. We next analyzed the ZPL positions of defects in the pre-oxidized samples, Figure 4b, and finding that they remained largely consistent with the standard growth samples as 87% of the 76 measured ZPLs are localized between 600-650 nm. We interpret this result as suggesting that increasing the crystallinity of the films alone, by decreasing the nucleation density and resulting growth kinetics, is not sufficient to alter the preferential formation of the structural defect responsible for emission from 600-650 nm. As both standard and pre-oxidized conditions experience similar B diffusion into Cu, and thus similar Cu phase and B feeding rates, only the number of defects observed is altered not the energetics of which defects are created.

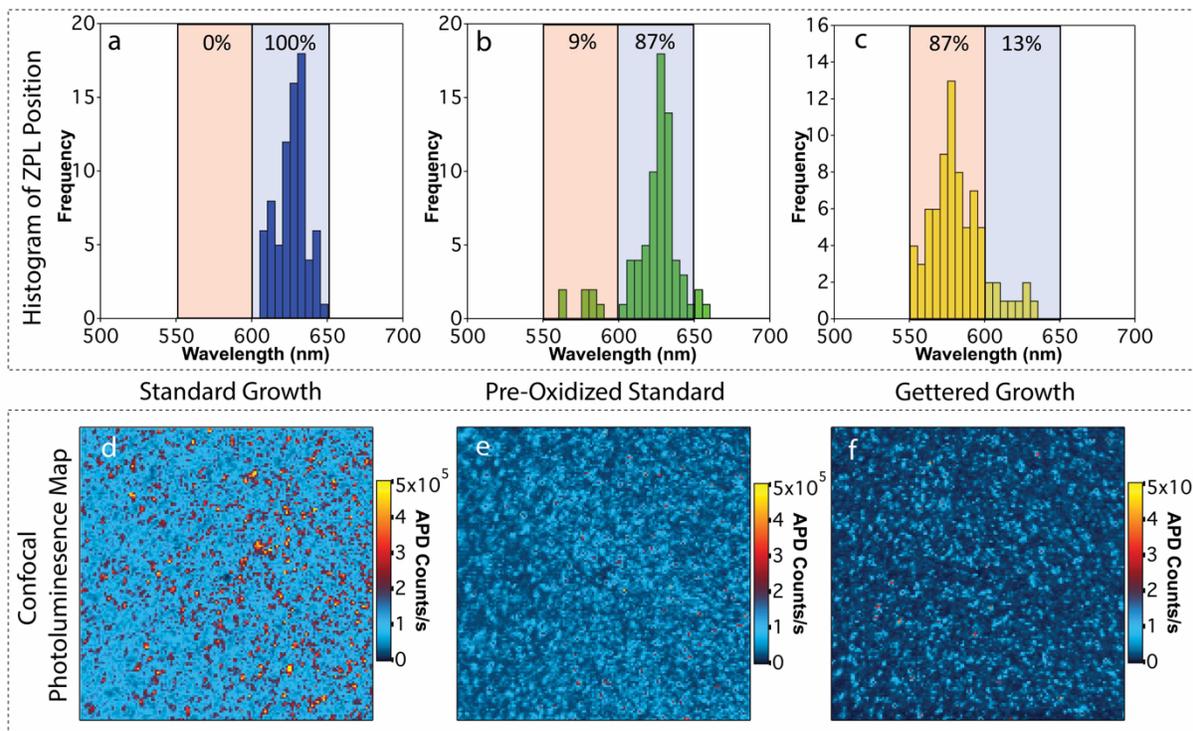

*Figure 4. Control over the ZPL energy of hBN SPEs. (a-c) ZPL histograms showing the distribution of 76 hBN SPEs found under different growth conditions. (a) Standard growth samples show 100% of characterized emitters are found from 600-650 nm. (b) Pre-oxidized samples show a distribution of emitters still primarily from 600-650 nm (87%) suggesting that decreased nucleation density, and increased film crystallinity alone does not modify ZPL location. (c) Gettered growth samples display a distribution of emitters now located between 550-600 nm (87%). This suggests that the ZPL movement is due to the gettering effect. (d-f) Confocal PL maps taken from each growth type, with excitation by a 200μW 532 nm source using a 0.9NA objective. (d) Standard growth samples display a high density of luminescent point defects. (e) Pre-oxidized samples show a decrease in overall luminescence of the material, as well as the population of point defect like emission. (f) Gettered growth samples showing the lowest population of defects and overall luminescence of the hBN film.*

In contrast, the Ni backed gettered samples, Figure 4c, showed a dramatic change in the observed ZPL position of the emitters, with 87% of 76 investigated emitters now being located in the spectral range from 550-600 nm, Figure 4c. By mitigating the boron backside diffusion we can preferentially select for the formation of an alternative structural defect species populating the spectral region from 550-600 nm. These results also hint that boron vacancies may be involved in SPEs emitting in the spectral range of 550-600 nm, and substitutional boron defects may play a role in emitters in the 600-650 nm range. While both the pre-oxidized and gettered samples increase the crystallinity of the top side film growth, only the gettered samples show a shift in the ZPL histogram, definitively isolating the cause of this spectral shift to changes in the B diffusion into Cu. Note all other growth conditions are identical between the standard, pre-oxidized, and gettered samples other than the Cu substrate preparation. We also believe that the presence of Ni in the growth chamber can be ruled out as the cause of the shift, as previous studies have found that the same family of hBN defects were incorporated during growth on Cu and Ni, when all other parameters are held the same.[24] Additionally we can rule out other potential causes, such as sample thickness, precursor purity, copper foil batch, or transfer technique, as they were similarly held consistent. To the best of our knowledge, this constitutes the first systematic demonstration of targeting two distinct spectral regions for SPEs in hBN, providing a template for defect engineering of hBN SPEs across the entire observed ZPL distribution based on controlling catalyst bulk reservoir effects.[33]

Prior studies utilizing top down techniques, such as plasma processing,[32] strain engineering,[31] and electron-beam processing[50, 51] have found weak correlations between the method used and the spatial location of emitters created, however, offer no control over the ZPL energy distribution. Additionally, while bottom up techniques have previously been shown superior to top-down methods in terms of isolating a narrow region of emission energies, this localization has only been achieved for SPEs operating in the 550-600 nm range.[24, 25] By controlling the concentration of boron species supplied from the Cu catalyst, we can deterministically select the observed localized energy region to be positioned either from 550-600 nm or from 600-650 nm. Through further modification of the CVD growth conditions, the entire range of hBN SPEs energies, ~1.6-2.4eV, should be possible to realize with such techniques. Our results are also significant in segregating the previously uniform distribution of emitters ZPL energies across the 800 meV range into at least three defined energy distributions, 550-600 nm, 600-650 nm, and 650-800 nm each likely representing a different structural defect. Further studies into the photophysical properties of each ZPL region, may finally permit the structural determination of defects responsible for each emission window.

Interestingly, the region of ZPL localization for our standard growth samples is in contrast to that observed in LPCVD growth studies, where the bulk of the emission ZPLs are reported to occupy the region of 550-600 nm,[24, 25] suggesting the standard conditions favor the formation of a different structural defect than found in previous studies. Note the only substantial change to the growth conditions is operation under atmospheric pressure as opposed to low pressure, however this has been known to result is substantial changes to the resulting hBN films. While the structural nature of the defect(s) in hBN are still largely debatable within the field,[52] the most straightforward interpretation of ZPL localization in discrete energy regions is the preferential incorporation of different structural defects, one with ZPL positions from ~550-600 nm and another with ZPL positions from ~600-650nm. While alternate explanations involving variations in the local strain environments, or particular charge states of the same structural defect can be invoked and not definitively ruled out, the apparent homogeneity of strain or charge state required to explain the localization into two different regions appears unlikely. To date strain induced spectral shifts have only demonstrated maximum shift magnitudes of 2-3 nm making this an unlikely explanation for the observed spectral shift.[19] Conversely, while the potential of varying charge states of the same structural defect may be more suitable to explain the results based on demonstrations of spectral jumping of up to 100 nm, this has been previously attributed to photochemical reactions taking place at the flake surface and/or trapped carrier induced stark shifts,[53] neither of which appear suitable to explain the consistency of the localization observed in our study across multiple growths and samples. Furthermore, these conclusions are supported by recent photophysical characterization of SPEs within these specific spectral regions.[54]

We have also investigated the density of SPE incorporation across the three sample types, and are able to correlate this with the observed crystallinity. The normal growth conditions yield the highest density of SPE incorporation and sample background luminescence, suggesting the incorporation of both SPE and non-SPE defects at a higher rate, Figure 4d. In the pre-oxidized samples, which have an intermediate crystallinity, we find that there is a decrease in the sample background luminescence, suggesting fewer SPE and non-SPE defects, Figure 4e. This is consistent with the increasing single crystal domain size, clearly observable in the optical images seen in Figure 1f. For the Ni backed gettered samples, we find the lowest background luminescence but a density of SPEs roughly equal to the pre-oxidized growth, Figure 4f, confirming that the gettering technique is more effective than pre-growth oxidation alone for the preparation of hBN for nanophotonics applications. This constitutes the first controlled approach to select the observed density of SPEs hosted within hBN, which is desirable for integration into hybrid quantum systems such as cavities and waveguides which often require a single isolated emission center coupled to the structure. All three sample types also display extremely low background counts, 3-10 kcps, a significant improvement over other hBN material sources which can display high background count rates, further bolstering the potential of CVD grown hBN.[19]

**Conclusion**

In summary, we have adapted a gettering based approach to the APCVD growth of hBN, improving the crystallinity of the resulting hBN films over existing pre-oxidation methods. More importantly, we show that the gettering approach allows for the deterministic selection of SPEs with a particular emission energy, at a probability of greater that >85%, providing a template for the rational incorporation of hBN SPEs with desired properties. This also hints that SPE defects with ZPL energies in the region 550-600 nm are of a different structural nature than those from 600-650 nm, and likely suggesting that those >650 nm are the result of yet another structural defect. We also leverage the increasing crystallinity across the three as-grown sample types to demonstrate control the density of incorporated SPEs, while simultaneously achieving extremely low background fluorescence from CVD samples. The ability to select ZPL energies and emitter densities makes bottom up fabrication techniques a leading solution for quantum photonics based on hBN.

**Methods/Experimental**

*Chemical vapor deposition of few-layer hBN films on copper substrates:*

The h-BN was obtained on the Cu foil via APCVD growth, using ammonia borane powder (AB) as a solid precursor. A piece of 2 cm x 4 cm of 25 μm thick Cu foil (Alfa Aesar, 99.8%) was chemically polished in the acetic acid bath for 5 minutes followed by rinsing with water to remove all the contamination adsorbed on the surface. The Cu substrate placed on quartz/nickel support substrate and loaded into the quartz tube furnace. The reaction chamber was flushed with a 500 sccm flow of Argon (Ar) for 20 minutes to remove entrapped air. The furnace heated to the growth temperature of 1040°C under the flow of Ar/H2 (20:1) mixture and annealed for 60 minutes, for "standard" growth conditions. For pre-oxidized conditions, after cleaning the Cu foil, it was prepared by heating at 200°C for 5 minutes in air to obtain copper oxide layer on Cu surface prior to growth. Furnace ramping was then performed under a flow of pure Ar to preserve the oxide layer. Thereafter, h-BN growth was commenced by heating the AB powder at 100°C to introduce the precursor into the reaction chamber with flow of carrier gas mixture (Ar/H2). The growth continued for certain time duration (5, 10 and 20 minutes) to obtain full coverage of the h-BN film. Finally, the growth terminated by taking the samples out of heating zone and cooled down to the room temperature under a 500 sccm flow of Ar. The detailed growth profile is illustrated in the schematic shown in Figure S2.

*Transfer of hBN films:*

The h-BN films were transferred to 300 nm $SiO_2$/Si substrate using polymer assisted wet transfer method. The 300 nm of PMMA were spin coating (3000 rpm) on the as-grown h-BN films on Cu foil as a mechanical support for transfer. The stacks were baked at 180°C for 5 minutes. The PMMA/h-BN stack were detached by etching the Cu into $FeCl_3$ aqueous solution. The PMMA/h-BN stack were than cleaned using dilute HCl for half an hour. Than the samples were scooped up using desired substrate (300 nm $SiO_2$/Si), and dried at 70°C for an hour. Subsequently PMMA is removed by acetone vapors and samples were blown dried using N2 stream.

*Optical studies of hBN single photon emitters:*

PL studies were carried out using a home-built scanning confocal microscopy with continuous wave (CW) 532-nm laser (Gem 532, Laser Quantum Ltd.) as excitation. The laser was directed through a 532 nm line filter and a half-waveplate and focused onto the sample using a high numerical aperture (100x, NA = 0.9, Nikon) objective lens. Scanning was performed using an X−Y piezo fast steering mirror (FSM-300). The collected light was filtered using a 532-nm dichroic mirror (532 nm laser BrightLine, Semrock) and an additional long pass 568-nm filter

(Semrock). The signal was then coupled into a graded-index multimode fiber, where the fiber aperture of 62.5 μm serves as a confocal pinhole. A flipping mirror was used to direct the emission to a spectrometer (Acton Spectra Pro, Princeton Instrument Inc.) or to two avalanche photodiodes (Excelitas Technologies) in a Hanbury Brown-Twiss configuration, for collection of spectra and photon counting, respectively. Correlation measurements were carried out using a time-correlated single photon counting module (PicoHarp 300, PicoQuant). All the second-order autocorrelation $g^2(\tau)$ measurements were analyzed and fitted without background correction unless otherwise specified.

**Acknowledgement**

This project was supported by the Research Grant Council of Hong Kong SAR (Project number 16204815), NSFC-RGC Joint Research Scheme (N_HKUST607/17), the Innovation and Technology Commission (ITC-CNERC14SC01), the Guangzhou Science & Technology (Project 201704030134). Technical assistance from the Materials Characterization and Preparation Facilities, HKUST, is greatly appreciated. Financial support from the Australian Research council (via DP180100077, DP190101058), the Asian Office of Aerospace Research and Development grant FA2386-17-1-4064, the Office of Naval Research Global under grant number N62909-18-1-2025 are gratefully acknowledged. This research is supported by an Australian Government Research Training Program Scholarship.

**Supporting Information**

Figures S1-S7: Information describing methods, setups, and fittings utilized. Additional results on the growth process, AFM traces of the transferred flakes, and peak fitting of Raman spectra. This material is available free of charge *via* the Internet at http://pubs.acs.org.